\documentclass[aps,10pt,twocolumn,superscriptaddress]{revtex4-1}

\usepackage{footnote}
\usepackage{times}
\usepackage{booktabs}
\usepackage{bm,natbib}
\usepackage{inputenc}
\usepackage{graphicx}
\usepackage{mathrsfs}
\usepackage{dcolumn,fancyhdr}
\usepackage{amsmath}
\usepackage{amssymb}
\usepackage{amsfonts,gensymb}
\usepackage{indentfirst}
\usepackage{bbold}
\usepackage{multirow}
\usepackage{dsfont}
\usepackage[usenames,dvipsnames]{xcolor}      

\usepackage{changes}

\usepackage{hyperref}
\hypersetup{
colorlinks=true,
linkcolor=blue,
citecolor=blue,
filecolor=magenta,
urlcolor=cyan
}

\usepackage[capitalize]{cleveref}

\newcommand*{\Me}[1]{\textcolor{black}{#1}}

\newcommand{\tr}[1]{\mathrm{Tr}\left[ {#1} \right]} 
\newcommand{\Tr}[2]{\mathrm{Tr}_{#1}\left[ {#2} \right]}
\newcommand{\ket}[1]{\vert #1\rangle}
\newcommand{\bra}[1]{\langle #1\vert}

\newcommand*\dagg{^{\dagger}}	
\newcommand{\matr}[1]{\mathsf{#1}}
\newcommand{\id}{1}
\newcommand*\mat[1]{\begin{pmatrix}#1\end{pmatrix}}

\begin{document}
\title{Stroboscopic quantum optomechanics}
\author{Matteo Brunelli}
\affiliation{Cavendish Laboratory, University of Cambridge, Cambridge CB3 0HE, United Kingdom}
\author{Daniel Malz}
\affiliation{Max-Planck-Institut f\"ur Quantenoptik, Hans-Kopfermann-Strasse 1, D-85748 Garching, Germany}
\author{Albert Schliesser}
\affiliation{Niels Bohr Institute, University of Copenhagen, 2100 Copenhagen, Denmark}
\affiliation{Center for Hybrid Quantum Networks (Hy-Q), Niels Bohr Institute, University of Copenhagen, 2100 Copenhagen, Denmark}
\author{Andreas Nunnenkamp}
\affiliation{Cavendish Laboratory, University of Cambridge, Cambridge CB3 0HE, United Kingdom}

\date{\today}

\begin{abstract}
We consider an optomechanical cavity that is driven stroboscopically by a train of short pulses. By suitably choosing the inter-pulse spacing we show that ground-state cooling and mechanical squeezing can be achieved, even in the presence of mechanical dissipation and for moderate radiation-pressure interaction. We provide a full quantum-mechanical treatment of stroboscopic backaction-evading measurements, for which we give a simple analytic insight, and discuss preparation and verification of squeezed mechanical states. We further consider stroboscopic driving of a pair of non-interacting mechanical resonators coupled to a common cavity field, and show that they can be simultaneously cooled and entangled. Stroboscopic quantum optomechanics extends measurement-based quantum control of mechanical systems beyond the good-cavity limit. 

\end{abstract} 
 
\maketitle

\section{Introduction}
Cavity optomechanics has proven extremely successful in controlling nanoscale and microscale mechanical motion at the quantum level~\cite{OptoRev}.
Among the key achievements is the demonstration of ground state cooling~\cite{Chan12,Teufel11}, mechanical squeezing~\cite{Wollman15,Lecocq15,Pirkkalainen15} and mechanical entanglement~\cite{MechEnt18}. Most of these milestones have been obtained in sideband-resolved optomechanical systems operating in the continuous-wave or amplitude-modulated (two-tone) regime, where a notion of stationary regime can be defined, in some suitable rotating frame. Going beyond steady-state operation may be beneficial for several reasons, e.g.~it allows to circumvent stability requirements. Sideband-resolved optomechanical systems driven by long pulses have been considered both for controlling mechanical motion~\cite{Hofer11,Machnes12,Liao11,Vostrosablin16} and as a model of quantum interface between flying quantum carriers; for instance, entanglement between microwave and mechanical degrees of freedom~\cite{Palomaki13Ent} and quantum state transfer~\cite{Palomaki13Transfer} have been demonstrated in this regime.

Pulsed protocols can also lift the stringent requirement of sideband resolution. By employing pulses much shorter than the mechanical period, quantum state  preparation and readout, e.g. of low-entropy and squeezed mechanical states~\cite{VannerPulsed,Bennett16,Hoff16,Bennett18,Vostrosablin18,Khosla17,Asjad14}, as well as opto-mechanical  and all-mechanical entanglement~\cite{Clarke19}  can in principle be achieved. However, in order to neglect non-unitary processes, coherent operations are restricted to very short times (also less than a single mechanical cycle). The conditional preparation of quantum states with few pulses also requires large interaction strengths, which has so far prevented pulsed optomechanics to attain the quantum regime~\cite{Vanner13Exp}. Only very recently, pulsed operation close to the quantum regime has been demonstrated in a setup based on a photonic crystal nanobeam~\cite{Muhonen19}. 
 
In this work we take a different approach and study the conditional dynamics of an optomechanical system driven by a train of pulses. We show that this new regime---{\it stroboscopic quantum optomechanics}---is effective to prepare and verify quantum states of mechanical motion beyond the sideband-resolved regime. In particular, by suitably choosing the spacing between the pulses, ground state cooling and squeezing of a single mechanical resonator can be achieved, as well as collective cooling and entanglement of two non-degenerate resonators (radiation-pressure coupled to a common cavity mode). 

Compared to single-pulse protocols, our approach has the distinct advantage to allow for a cumulative effect of the measurements over many mechanical cycles, thus relaxing the requirement on the optomechanical coupling strength considerably. This however requires including mechanical dissipation in the description of the dynamics, as opposed to Refs.~\cite{Bennett16,Hoff16,Bennett18,Vostrosablin18,Khosla17}. Due to the competition between radiation-pressure interaction and mechanical dissipation, the mechanical system eventually settles into a steady state, albeit a periodic one. For such a \emph{stroboscopic steady state}, we provide simple analytic expressions for the conditional state. From this point of view, our work draws an interesting connection between the conditional dynamics of periodically measured systems and the Floquet theory of optomechanics~\cite{Malz16,Qiu19}. 

Our study is inspired by early works in backaction-evading (BAE) measurements,  where the stroboscopic dynamics of mechanical transducers was studied for the detection of weak classical signals~\cite{Braginsky2,Marchese92,Onofrio90,Vasilakis11,BAESpin15}. We provide a full quantum-mechanical treatment of stroboscopic BAE measurements~\cite{WeakMeasRev,OnofrioRev}, which was so far missing. We discuss in details corrections to the ideal measurement regime stemming from thermal decoherence and the finite length of each pulse. Notably, we show that including the latter effect, usually considered detrimental, enables preparing  \emph{pure} mechanical squeezed states and opto-mechanical entanglement. In short, we show that stroboscopic quantum optomechanics bypasses the need for strong measurements and provides an effective and versatile tool for measurement-based quantum control of mechanical states.

The rest of the paper is organized as follows: in Sec.~\ref{s:Model} we describe the system and derive an effective model of the dynamics based on stroboscopic measurements. The predictions of this model for stroboscopic squeezing and cooling of mechanical motion are presented in Sec.~\ref{s:Squeez} and Sec.~\ref{s:Cool}, respectively. In Sec.~\ref{s:Improved} we discuss engineering squeezed quantum states in connection with stroboscopic BAE measurements of mechanical motion. In Sec.~\ref{s:Verification} we implement verification of the conditional state via retrodiction. In Sec.~\ref{s:TwoMode} we extend stroboscopic quantum optomechanics to non-interacting mechanical resonators coupled to a common cavity field, and show that they can be simultaneously cooled and entangled. In Sec.~\ref{s:Exp} we discuss some experimentally relevant considerations for implementing our ideas. Finally, Sec.~\ref{s:Conclusions} collects conclusive remarks and provides an outlook.

\section{A simple model of stroboscopic conditional dynamics} \label{s:Model}

\begin{figure}[t]
\centering
\includegraphics[width=\linewidth]{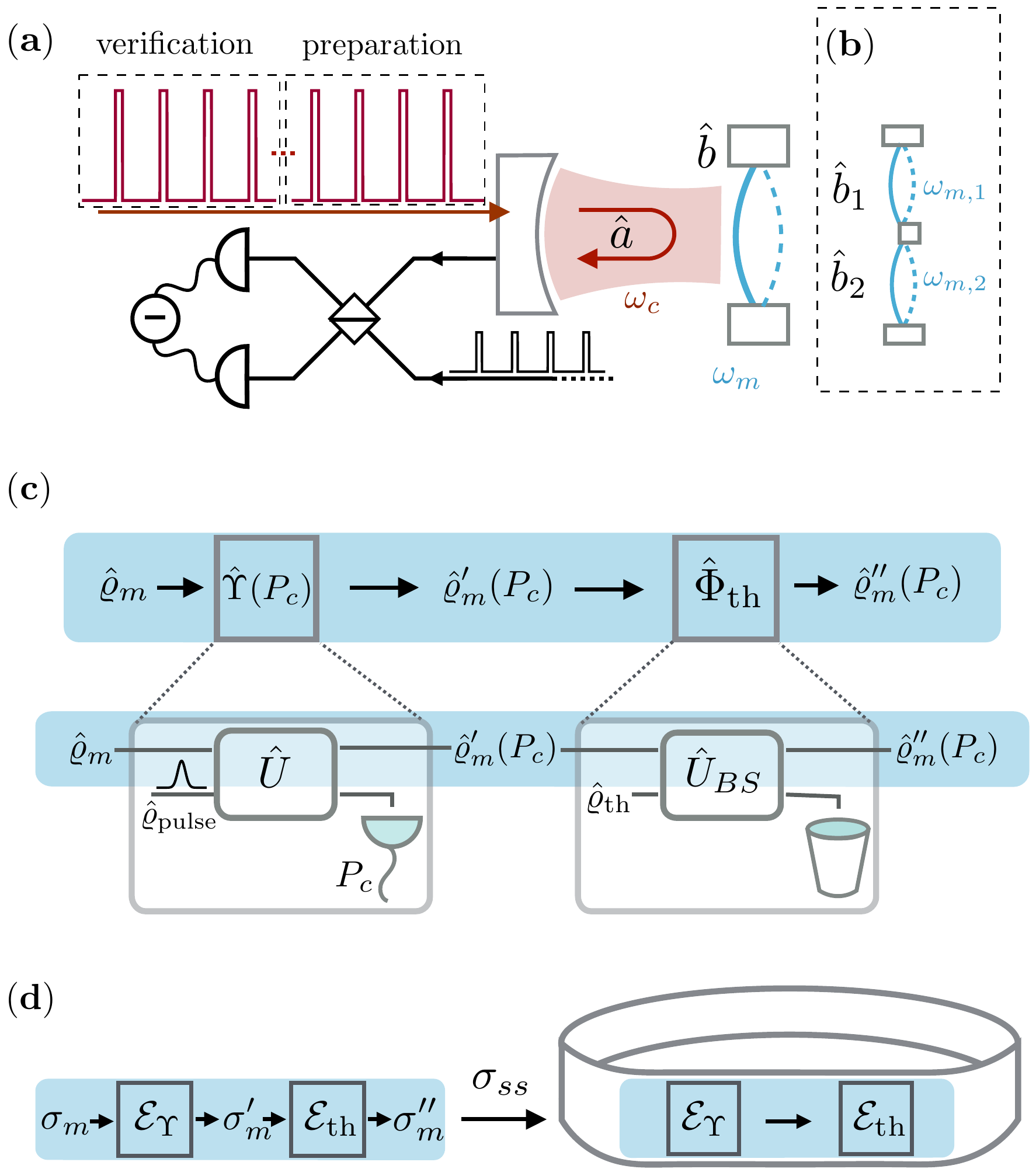}
\caption{
An optomechanical cavity ($\hat a$) is driven by a train of short pulses. After having interacted with the mechanical resonator ($\hat b$), the pulses are measured in reflection. Suitably choosing the spacing between the pulses, the setup implements either a stroboscopic backaction evading (BAE) measurement or measurement cooling. Mechanical squeezing and ground state cooling can be respectively achieved, conditional on the measurement record. A second train of pulses is used to verify the conditional state. ({\bf b}) Considering two mechanical modes $\hat b_1$, $\hat b_2$ instead, mechanical entanglement and collective cooling can be obtained. ({\bf c}) Elementary sequence of the stroboscopic protocol: an optical pulse, modeled by a generalized position measurement, is followed by free evolution and mechanical dissipation (see Sec.~\ref{s:Model} for details). For Gaussian input states both measurement and dissipation induce a deterministic transformation of the mechanical covariance matrix ({\bf d}). Repeating this  sequence yields a stroboscopic steady state, which is invariant under the joint action of measurement and dissipation. 
\label{f:Sketch}}
\end{figure} 

We consider a standard optomechanical system where the position $\hat x$ of a mechanical oscillator of frequency $\omega_m$ modulates the frequency of a cavity mode $\hat a$ of linewidth $\kappa$~\cite{OptoRev}. The cavity is illuminated with a train of short coherent pulses of length $\tau$ much smaller than the mechanical period, i.e. $\omega_m \tau\ll1$. The number of photons  $N_\mathrm{p}$ in each pulse is large enough to make linearization of the optomechanical interaction an excellent approximation, such that we have  ($\hbar=1$)
\begin{align}
\hat H_I(t)&=-g(t) \hat X_c \left(\hat X_m \cos\omega_m t+\hat P_m \sin\omega_m t \right) \,, \label{QND2}
\end{align}
where $g$ is the pulsed coupling constant,  $\hat X_c=(\hat a+\hat a^\dagger)/\sqrt2$ and we expressed the mechanical position in terms of the slowly varying quadratures $\hat X_m,\,\hat P_m$. During the interaction time the coupling induces  the unitary evolution
\begin{equation}\label{U}
\hat U(t,t_0)=\mathcal{T}\exp\left\{-i\int_{t_0}^{t} \mathrm{d}t_1\, \hat H_I(t_1)\right\}\,.
\end{equation}
For a very short pulse the harmonic motion can be neglected and the unitary evolution can be approximated as
\begin{equation}\label{Simple}
\hat U\approx e^{i\chi \hat X_m \hat X_c} \, ,
\end{equation} 
where $\chi$ quantifies the strength of the interaction. An estimate of the latter for a fast cavity in the adiabatic limit $\kappa\gg\tau^{-1}$ yields $\chi=2 g_0 \sqrt{\frac{N_\mathrm{p}\tau}{\kappa}}$, where $g_0$ is the single-photon optomechanical coupling. As we show in Appendix~\ref{app:Derivation}, in this limit the interaction is {\it de facto} instantaneous, in that mixing between the two mechanical quadratures is fully neglected. Equation~\eqref{Simple} realizes a quantum non-demolition (QND) gate between the optical and mechanical amplitudes~\cite{GaussRev}:  $\hat X_c$, $\hat X_m$ are left untouched and information about them is acquired by the conjugate quadratures $\hat U^\dagger \hat P_{m (c)} \hat U=\hat P_{m (c)} +\chi \hat X_{c (m)}$. The probability of recording a value $P_c$ of the optical phase quadrature after such interaction is given by 
\begin{align}
\Pr(P_c)&=\tr{\ket{P_c}\bra{P_c}\otimes \mathbb{1}_m \hat U \ket{0}\bra{0}\otimes \hat{\varrho}_m \hat U^\dagger}\label{Born1} \\
&=\Tr{m}{\hat \Upsilon^\dagger \hat \Upsilon \hat{\varrho}_m} \, , \nonumber
\end{align}
where the cavity starts off in the vacuum and the resonator in an arbitrary state $\hat{\varrho}_m$. In the second line of Eq.~\eqref{Born1} we have rewritten the probability by introducing the family of Kraus operators $\hat \Upsilon(P_c)=\bra{P_c}\hat U\ket{0}$, elements of the positive operator-valued measure (POVM) $\{\hat \Upsilon^\dagger \hat \Upsilon \}_{P_c}$, that satisfy $\hat \Upsilon^\dagger \hat \Upsilon\ge0$, $\nobreak{\int \mathrm{d}P_c \hat \Upsilon^\dagger \hat \Upsilon=\mathbb{1}_m}$. An explicit expression for $\hat \Upsilon$ is given by
\begin{align}
\hat \Upsilon( P_c)
&=\tfrac{1}{\pi^{1/4}}\, e^{-\frac12 \left( P_c - \chi\hat X_m\right)^2} \, , \label{Kraus}
\end{align}
which shows that the effect of the pulse on the mechanics is akin to a generalized position measurement. This expression has been first used to model momentum diffusion in continuous weak measurements~\cite{Caves87}. Later, it was employed  to model an optomechanical system driven by a single strong pulse,  i.e., the regime of pulsed quantum optomechanics~\cite{VannerPulsed}. From Eq.~\eqref{Kraus} we notice that, when acting on a pure state, the measurement operator multiplies the wave function by a Gaussian function of width $\chi^{-2}$ and centered around the position $P_c/\chi$; by increasing the  interaction strength $\chi$, the wave function thus gets increasingly localized in position. Upon recording the outcome $P_c$, the mechanical density matrix is transformed as
\begin{equation}\label{Cond}
\hat{\varrho}_m\rightarrow\tfrac{1}{\Pr(P_c)} \hat \Upsilon(P_c) \hat{\varrho}_m \hat \Upsilon(P_c)^\dagger\,,
\end{equation} 
which is the conditional, or post-measurement state.

When the pulse is off, environment-induced decoherence is affecting the otherwise free evolution of the mechanical resonator. The dynamics is governed by
\begin{equation}\label{Diss}
\mathcal{L}\hat \varrho_m=-i[\hat H_0,\hat \varrho_m ]+\gamma (\bar n+1)\mathcal{D}[\hat b]\hat  \varrho_m+\gamma \bar n\mathcal{D}[\hat b^\dagger]\hat  \varrho_m
\end{equation}
where $\bar n$ and $\gamma$ are the mean occupation and damping rate of the mechanical bath and $\hat b$ is the annihilation operator associated to mechanical quadratures. The evolution over a finite amount of time is given by the map $\hat \Phi_\mathrm{th}=e^{\mathcal{L}t}$. The pulsed interaction~\eqref{Cond} and the free-evolution-plus-dissipation~\eqref{Diss} form the `unit cell' of our stroboscopic model, which can be thought of as a repetition of these two elementary steps, see Fig.~\ref{f:Sketch}({\bf c}). 
As we discuss below, when concatenating many such steps one is free to choose the spacing between two subsequent pulses. Over this time (i) the mechanical mode picks up a phase, which determines which quadrature is measured at the next interaction, and (ii) the mechanics exchanges phonons with the thermal environment. In particular, the presence of the latter contribution---neglected in previous studies~\cite{Bennett16,Hoff16,Bennett18,Vostrosablin18,Khosla17}---competes with the measurement, eventually leading to a non-equilibrium steady state.

\subsection{From measurement-induced evolution to deterministic CP maps}\label{s:CPMap}
A great simplification comes from assuming that both the measurement and the dissipation act on a Gaussian state, in which case their output is a Gaussian state too~\cite{GaussRev,Olivares12,Marco16}. 
For the case of a Gaussian measurement, such as the quadrature measurement in Eq.~\eqref{Born1}, the post-measurement state~\eqref{Cond} depends on the measurement outcome only through the first moments or, equivalently, the measurement-induced evolution of the second moments is deterministic; this is a general feature of Gaussian measurements~\cite{AlessioBook}. Therefore, the effect of the measurement can be cast in the form of a deterministic map $\mathcal{E}_\Upsilon$ for the second statistical moments~\cite{Giedke02}. The action of  this map $\sigma'=\mathcal{E}_\Upsilon(\sigma)$ on the mechanical covariance matrix $\sigma$ (with variance $\sigma_{X_m},\, \sigma_{P_m}$ and covariance $\sigma_{X_mP_m}$) is given by
\begin{align} \label{PostMeasVar}
\sigma_{X_m}'&=\frac{\sigma_{X_m}}{1+2\chi^2 \sigma_{X_m}}\, ,\\
\sigma_{P_m}'&=\frac{\chi^2}{2} +\frac{\sigma_{P_m}+2\chi^2(\sigma_{X_m}\sigma_{P_m}-\sigma_{X_mP_m}^2)}{1+2\chi^2 \sigma_{X_m}}\, ,\\ 
\sigma_{X_mP_m}'&=\frac{\sigma_{X_mP_m}}{1+2\chi^2 \sigma_{X_m}}\, .
\end{align}
We explicitly see that the stochastic component of the measurement ($P_c$) is absent from the above expressions. The first and second expression describe the reduction of the variance along $\hat X_m$,  and the increased fluctuations of the conjugate quadrature due to the quantum backaction, respectively. 

The (commutative) action of dissipation and free evolution \eqref{Diss} on the covariance matrix is described by the map $\mathcal{E}_\mathrm{th,\phi}(\sigma)=e^{-\gamma t}\mathcal{R_\phi} \sigma \mathcal{R_\phi}^T+(1-e^{-\gamma t})\sigma_\mathrm{th}$, where $\sigma_\mathrm{th}=(\bar n+\tfrac12)\mathbb{1}_2$ is the covariance matrix of a thermal state and $\mathcal{R_\phi}$ is the rotation matrix due to harmonic evolution. Equivalently, under $\mathcal{E}_\mathrm{th,\phi}$, the input state gets rotated and mixed with a thermal state via a beam splitter of effective transmissivity $\eta=e^{-\gamma t}$.

Measurement and dissipation compete over time. The former tries to reduce the uncertainty in one quadrature (at the expense of the other), while the latter tries to restore isotropy. Crucially, the spacing between two pulses determines the amount of mixing between the quadratures from one measurement to the next one. This consideration applies to {\it any} sequence of equally spaced pulses; for example, one can obtain a recursion relation $\sigma^{(N)}=(\mathcal{E}_\Upsilon \circ \mathcal{E}_{\mathrm{th},\phi})\sigma^{(N-1)}$ to model a short train of pulses. This operation regime has recently become experimentally relevant for quantum applications~\cite{Muhonen19}. Here we focus on a different regime: when the action of the measurement is undone by the dissipation there is no net effect over a `unit cell' and the system reaches a {\it stroboscopic steady state} [see Fig.~\ref{f:Sketch}({\bf d})]. More formally, this state is a fixed point of the map $\mathcal{E}_\Upsilon \circ \mathcal{E}_\mathrm{th,\phi}$, namely it satisfies $\sigma_{ss}=(\mathcal{E}_\Upsilon \circ \mathcal{E}_\mathrm{th,\phi})\sigma_{ss}$. We stress that the two operations do not commute, so that in general $\mathcal{E}_\Upsilon \circ \mathcal{E}_{\mathrm{th},\phi}\neq\mathcal{E}_{\mathrm{th},\phi} \circ \mathcal{E}_\Upsilon$, as we shall see below. This is a novel regime for cavity optomechanics, which has focused either on steady state properties of continuously driven systems or in the finite-time dynamics, as in pulsed optomechanics.

\section{Stroboscopic squeezing of mechanical motion}\label{s:Squeez}
\begin{figure}[t!]
\centering
\includegraphics[width=\linewidth]{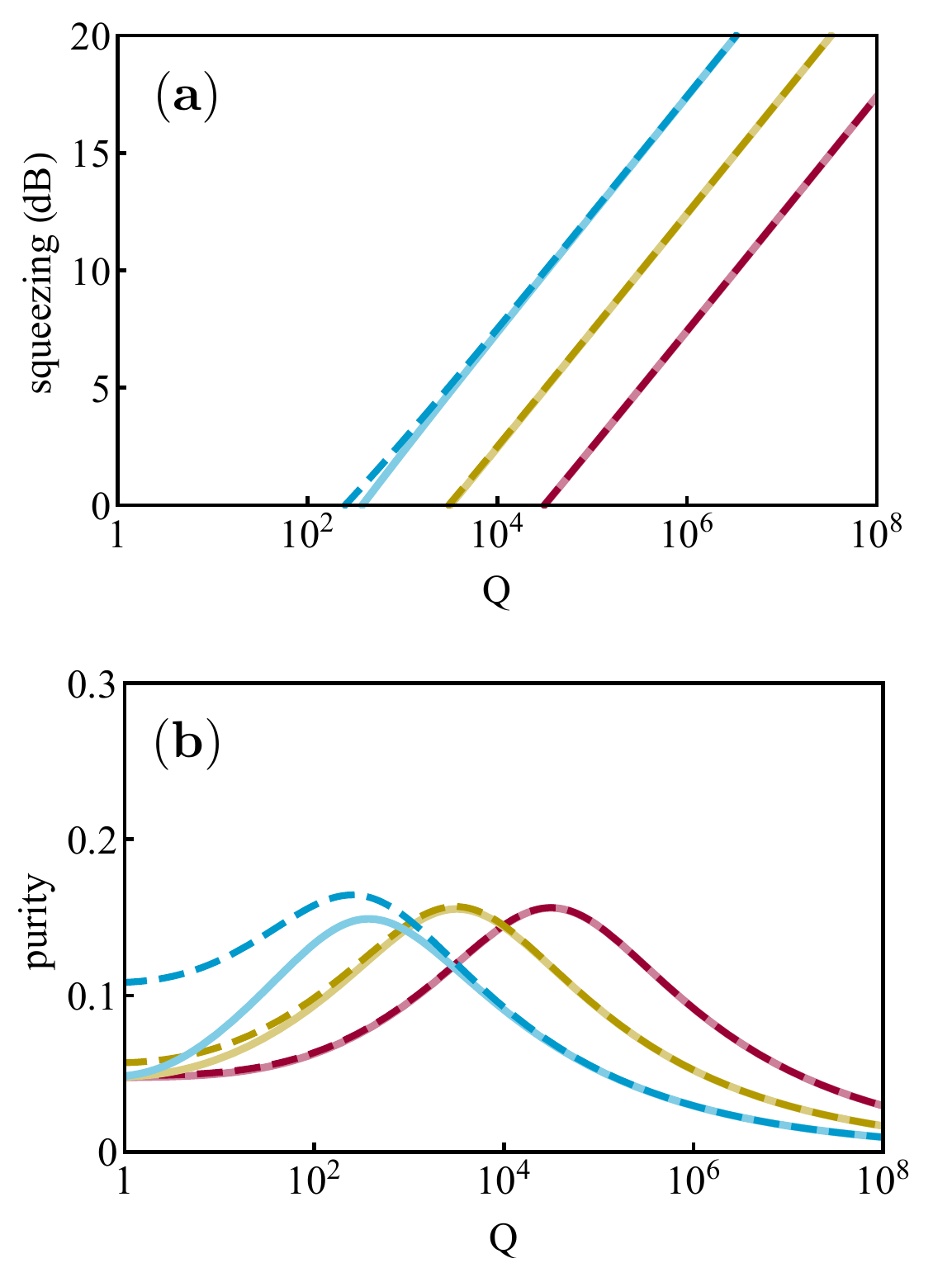}
\caption{Steady-state value of mechanical squeezing ({\bf a}) and mechanical purity ({\bf b}) for a stroboscopic BAE measurement. Values of the coupling are $\chi=0.05$ (red), $\chi=0.1$ (yellow), $\chi=0.5$ (cyan). Solid lines are for the stroboscopic steady state of the map $\mathcal{E}_\mathrm{th} \circ \mathcal{E}_\Upsilon$, while dashed \Me{ones for $\mathcal{E}_\Upsilon \circ \mathcal{E}_\mathrm{th}$}. The mean phonon number is set to $\bar n=10$.
\label{f:PlotStrob1}}
\end{figure}

\subsection{Stroboscopic BAE measurement}\label{s:StroboBAE}
The first case we consider is that of a \emph{stroboscopic BAE measurement}, for which a classical treatment is discussed in Refs.~\cite{Braginsky2,WeakMeasRev,OnofrioRev}. By choosing pulses interspaced by a multiple of half the mechanical period, we can in principle realize a QND measurement of position. Indeed,  one has $[\hat x(t),\hat x(t+T)]=i\sin(\omega_m T$), so that at the stroboscopic times $T=k \pi/\omega_m$ a sequence of precise position measurement is possible with no fundamental limit imposed by quantum mechanics (in the following we always take the shortest interval $k=1$). However, due to the presence of the environment, the covariance matrix does not come back to itself half a period later. We then look for solutions where the \emph{combined} action of the measurement and the environment leaves the state invariant. Solving for the stroboscopic steady state $\sigma_{ss}=(\mathcal{E}_\Upsilon \circ \mathcal{E}_{\mathrm{th},\pi})\sigma_{ss}$, we get
\begin{align}
\sigma_{X_m}&=\frac{2 \bar n+1}{1+z+\sqrt{1+z^2+2 z \coth \left(\frac{\gamma  T}{2}\right)}} \label{SqX} \,,\\
\sigma_{P_m}&=\bar n+\frac{1}{2} +\frac{\chi ^2}{2 \left(1-e^{-\gamma  T}\right)}\, \label{SqP}\,,
\end{align}
and $\sigma_{X_mP_m}\equiv 0$, where we set $z=(2 \bar n+1) \chi ^2$.  We stress that the knowledge of such state is \emph{conditioned} on the stream of measurement results. These expressions can be considerably simplified for large values of the mechanical quality factor $Q=\omega_m/\gamma$. The leading terms in the expansion are given by
\begin{equation}\label{SqApprox}
\sigma_{X_m}=\frac{\sqrt{2\pi( \bar n+1/2)}}{2  \chi \sqrt Q}\, ,\qquad \sigma_{P_m}=\bar n+\frac12 +\frac{Q \chi^2}{2\pi}\,.
\end{equation}
This simple result provides a quantum-mechanical treatment of stroboscopic BAE measurement and proves that mechanical decoherence does not preclude the occurrence of squeezing at long times. Indeed, uncertainty may fall below the zero-point value, which implies a squeezed state of the resonator. Mechanical squeezing [expressed in $-10\log_{10}(2\sigma_{X_m})$ Decibel (dB)] is plotted Fig.~\ref{f:PlotStrob1}({\bf a}). We stress that different $Q$ entail different characteristic times to approach the stroboscopic steady state.

The solid lines are for the steady state relative to $\mathcal{E}_{\mathrm{th},\pi} \circ \mathcal{E}_\Upsilon$, while the dashed for $\mathcal{E}_\Upsilon \circ \mathcal{E}_{\mathrm{th},\pi}$. Physically, they correspond to the knowledge of the conditional state directly \emph{after} or directly \emph{before} the measurement. We can see discrepancies arising due to the non-commutative character of the two maps for low $Q$ and large coupling values. In this parameter regime, if we start from a thermal state, the effects of measuring first are (partially) undone by the subsequent application of the thermal channel. On the other hand, by reversing the order (i.e., considering the map $\mathcal{E}_\Upsilon \circ \mathcal{E}_{\mathrm{th},\pi}$) $\mathcal{E}_{\mathrm{th},\pi}$ acts as the identity, so the first measurement retains more conditioning power. The difference between the two cases thus boils down to an extra pulse, which has significative impact for large $\chi$ and explains the larger amount of squeezing. However, already for moderately large quality factors the two predictions coincide. 

It is interesting to compare the condition for mechanical squeezing enforced by Eq.~\eqref{SqApprox} with that required by pulsed optomechanics, i.e. by applying a single pulse~\eqref{Kraus}. For a single pulse, values of the coupling $\chi>1$ are required to obtain squeezing (independently of $\bar n$), which has so far precluded reaching the quantum regime in pulsed optomechanics experiments. On the other hand, with stroboscopic driving approaching $\sigma_{X_m}<1/2$ only requires $\chi>\sqrt{2\pi(\bar n+1/2)/Q}$, which can be considerably less demanding for large quality factors. \Me{In terms of the  multiphoton quantum cooperativity $\mathcal{C}_q$~\cite{OptoRev}, the above requirement reads $\mathcal{C}_q>8\pi/(\kappa\omega_m\tau^2)$.}

Finally, if we compare how the two variances in~\eqref{SqApprox} scale with $Q$, it is clear that fluctuations increase faster in $\hat P_m$ than they are reduced along $\hat X_m$. This means that while getting squeezed, the resonator gets also heated up. This fact is highlighted in Fig.~\ref{f:PlotStrob1}({\bf b}) where the mechanical purity $\mu=\tr{\hat \varrho_{ss}^2}$ for the same cases of panel ({\bf a}) is shown. In the large $Q$ limit the purity takes the simple form 
\begin{equation}
\mu=\pi^{1/4}  \left(\frac{Q}{2\bar n+1}\right)^{1/4} \sqrt{\frac{\chi }{2 \pi \bar n+Q \chi ^2}}\,.
\end{equation} 
Larger values of squeezing are accompanied by low purity. We will see in Sec.~\ref{s:Improved} that this conclusion gets drastically modified by considering the imperfect QND regime determined by mechanical evolution during the pulse.

 \begin{figure}[t!]
\centering
\includegraphics[width=\linewidth]{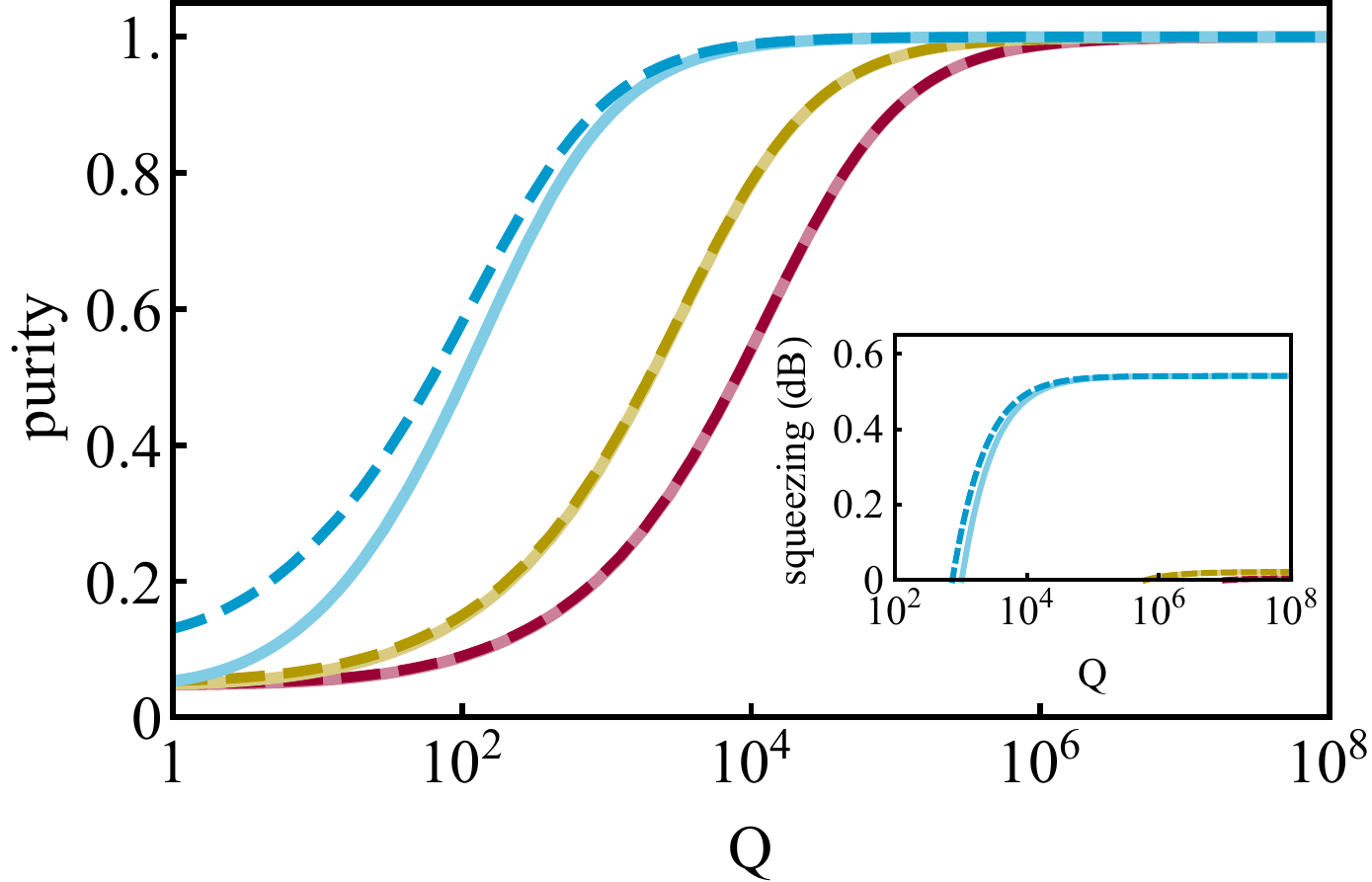}
\caption{Stroboscopic ground state cooling. Steady-state value of the mechanical purity for pulses spaced by a quarter of the mechanical period. Values of the coupling are $\chi=0.05$ (red), $\chi=0.1$ (yellow), $\chi=0.5$ (cyan). Solid lines are for the stroboscopic steady state relative to the map $\mathcal{E}_\mathrm{th} \circ \mathcal{E}_\Upsilon$, while dashed for the operations applied in reversed order. The mean phonon number is set to $\bar n=10$. In the inset we show residual asymmetry between the two quadratures [cf Eqs.~\eqref{SqXApproxCool},~\eqref{SqPApproxCool}], which can result in a small amount of squeezing.
\label{f:PlotStrob2}}
\end{figure}

\subsection{Squeezed input pulses}
Finally, we notice that the former results can be extended to the case where squeezed pulses, rather than coherent ones, are fed to the optomechanical cavity. In our simple model this observation amounts to replace the cavity vacuum seed state $\sigma_{\mathrm{pulse}}=\mathbb{1}/2$ [see Eq.~\eqref{Born1}] with  a squeezed state $\sigma_\mathrm{pulse}=\mathrm{diag}\bigl(\tfrac{e^r}{2},\tfrac{e^{-r}}{2}\bigr)$ squeezed along the  phase quadrature. This determines reduced fluctuations of the (measured) optical phase, which in turn enhances the conditioning effect of the measurement.  One obtains results as in Eqs.~\eqref{SqX}, \eqref{SqP} with the substitution $\chi\rightarrow e^\frac{r}{2}\chi$, namely a train of squeezed pulses magnifies the measurement strength by an exponential factor (in the degree of squeezing). 

\section{Stroboscopic ground state cooling}\label{s:Cool}

Another interesting case is obtained by spacing the pulses by a quarter of a period. In this case the value of the variance along $\hat X_m$ and $\hat P_m$ gets swapped by the free evolution, so that the measurement reduces both variances alternately.  An exact expression for the stroboscopic steady state is available also in this case, although quite cumbersome. For convenience below we give the expansion for large $Q$ 
\begin{align}
\sigma_{X_m}&=\frac{\sqrt{4+\chi^4}-\chi^2}{4}+\frac{\mathcal{F}(\chi,\bar n)}{Q} \,, \label{SqXApproxCool} \\
\sigma_{P_m}&=\frac{\sqrt{4+\chi^4}+\chi^2}{4}+\frac{\mathcal{G}(\chi,\bar n)}{Q}\,. \label{SqPApproxCool} 
\end{align}
The full expression of the functions $\mathcal{F}(\chi,\bar n)$, $\mathcal{G}(\chi,\bar n)$ is reported in Appendix~\ref{app:Cooling}. Unlike Eq.~\eqref{SqApprox}, now fluctuations in both quadratures converge  to a constant value for $Q\rightarrow \infty$. We also notice that there is a residual asymmetry between the two quadratures. The leading terms therefore describe a squeezed vacuum state, albeit one where the squeezing 
grows slowly with the coupling $\chi$.  For realistic values of the coupling the state is thus only weakly squeezed and has near-unit fidelity with the mechanical vacuum. The conditional purification of the mechanical state is also known as cooling-by-measurement~\cite{Vanner13Exp}. In the same spirit, we refer to this case as stroboscopic cooling.
Of course for finite values of the quality factor the steady state will be mixed, but cooling close to the ground state is still possible. We show these features in Fig.~\ref{f:PlotStrob2}. 

\section{Improved description and numerical simulations}\label{s:Improved}
In this Section we aim to provide a more accurate description of the stroboscopic conditional dynamics. We will focus on the case of stroboscopic BAE measurements but the analysis can be readily extended to the case of stroboscopic cooling. We expand along two directions: (i) we model the measurement as actually taking place outside the optical cavity and (ii) we evaluate the effects of the mechanical free evolution during the pulsed interaction. To this end, we consider the following extended Hamiltonian   
\begin{equation}\label{HExtended}
\hat H=\hat H_I(t) + i\sqrt\kappa (\hat a^\dag \hat a_{\mathrm{in},t}-\hat a\, \hat a_{\mathrm{in},t}^\mathrm{\dag}) \,,
\end{equation}
where, beside the term in Eq.~\eqref{QND2}, we also include an interaction with the continuum of electromagnetic modes $\hat a_{\mathrm{in},t}$  living outside the cavity.
This stream of modes interacts with the system at time $t$ and is otherwise uncorrelated $[\hat a_{\mathrm{in},t},\hat a_{\mathrm{in},t'}^\dag]=\delta(t-t')$. As customary, we assume they have Markovian correlation function $\langle\{\hat a_{\mathrm{in},t},\hat a_{\mathrm{in},t'}^\dag\}\rangle=\delta(t-t')$. 


For a short pulse of length $\tau$ (for now neglecting the free mechanical evolution) the corresponding propagator takes the form 
\begin{equation}\label{UExtended}
\hat U=e^{i\chi \hat X_m \hat X_c+i\sqrt{\kappa \tau}(\hat P_c \hat X_\mathrm{in}-\hat X_c \hat P_\mathrm{in})} \, ,
\end{equation}
where $\hat X_\mathrm{in},\,\hat P_\mathrm{in}$ are the proper (dimensionless) modes of the environment, i.e.,~$[\hat X_\mathrm{in},\hat P_\mathrm{in}]=i$, which are being measured; homodyne detection of the phase quadrature corresponds to projection along $\ket{P_\mathrm{in}}$ (see Appendix~\ref{app:ExtraMeas} for details).

\begin{figure}[t!]
\centering
\includegraphics[width=0.99\linewidth]{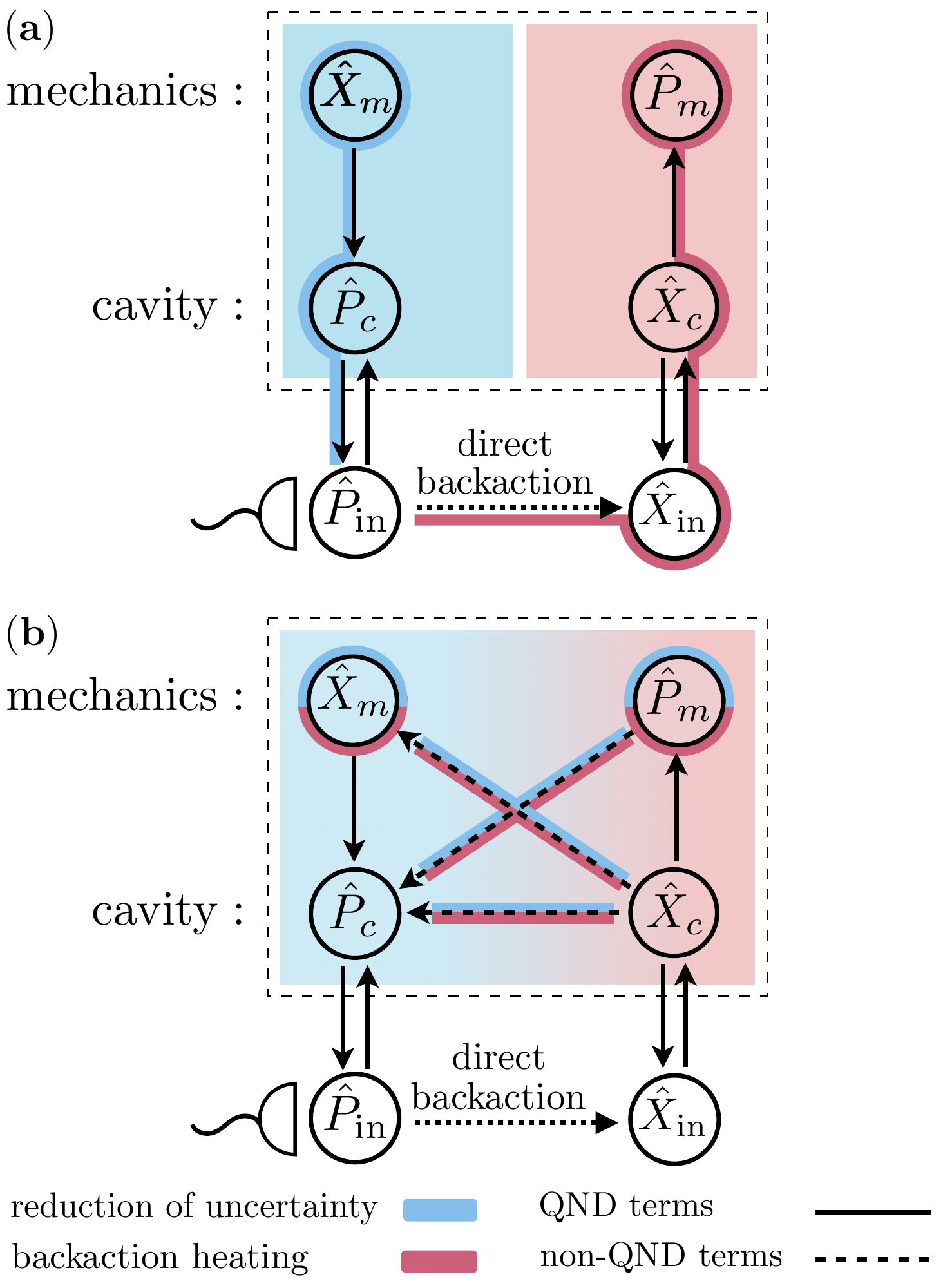}
\caption{Effects of the measurement for ({\bf a}) an ideal QND interaction and ({\bf b}) including mechanical free evolution during each pulse. The black arrows describe how quadratures influence each other in an ideal QND interaction [evolution according to Eq.~\eqref{UExtended}]. Measurement of the output phase quadrature both introduces backaction and allows to extract information. ({\bf a}) In an ideal stroboscopic measurement, measurement backaction and reduction of uncertainty take two distinct paths (the first is confined to $\hat P_m$ while the second to $\hat X_m$) which allows squeezing along $\hat X_m$. ({\bf b}) Non-QND terms open new paths (dashed arrows) where both conditioning and backaction spread. These reduce squeezing in $\hat X_m$, but at the same time enable acquiring information about all the quadratures, which results in larger mechanical purities and optomechanical entanglement.
\label{f:MeasScheme}}
\end{figure}

Formally, we can then proceed as in Sec.~\ref{s:Model} to compute the conditional covariance matrix of the optomechanical system, include thermal decoherence,  and enforce the stroboscopic steady-state condition.
The full expression of the conditional state of the mechanical system is quite cumbersome, but in the large $Q$ limit we get the following simple expressions
\begin{align}
\sigma_{X_m}&=\frac{\kappa \sqrt{2\pi( \bar n+1/2)}}{4 g \sin^2\left(\frac{\sqrt{\kappa  \tau }}{2}\right) \sqrt{\kappa\tau Q }}\,, \label{SqXextra} \\
\sigma_{P_m}&= \bar n+\frac{1}{2} +\frac{g^2 Q \tau [1- \cos \left(\sqrt{\kappa  \tau }\right)]}{\pi  \kappa } \label{SqPextra} \,,
\end{align}
with $g=2 g_0 \sqrt{\frac{N_\mathrm{p}}{\kappa\tau}}$. These expressions are to be seen as a refinement of Eq.~\eqref{SqApprox}; as we will show, they offer a useful comparison with numerical simulations.

Second, we include corrections to the ideal QND limit stemming from the mixing of the mechanical quadratures during a  pulse of finite length. The ensuing unitary evolution contains two new terms (see Appendix~\ref{app:Derivation} for the full expression): a squeezing term in the cavity amplitude, which however is $\mathcal{O}(g^2/\omega_m^2)$, and a spurious term $\propto \hat X_c \hat P_m$, which spoils the QND nature of the interaction. The strength of this term is $2\omega_m/\kappa$ times the QND part, so that the QND limit is approximately recovered only for optomechanical systems deep in the bad-cavity regime. It is therefore important to address the corrections arising for finite values of the sideband parameter, which limit the amount of conditional squeezing attainable. 

Due to the presence of quadrature mixing, a closed expression of the conditional state can no longer be found. However, we can get a clear physical picture of the effects brought about by non-QND term in the following way. 
Consider the effective Hamiltonian generating the optomechanical evolution, first neglecting and then including the non-QND term (to faithfully model the measurement, we also include the interaction with the extra-cavity modes); the corresponding expressions are given by Eq.~\eqref{QND2} and Eq.~\eqref{UnonQND}, respectively. 
\Me{We can use them to compute the Heisenberg evolution of the quadratures in both cases, $\dot{\hat{X}}_m=\ldots, \dot{\hat{P}}_m=\ldots$, and so on, where the terms appearing on the right-hand side drive the evolution of a given quadrature.}
The equations of motion for the two cases are schematized in Fig.~\ref{f:MeasScheme}({\bf a}) and ({\bf b}), respectively, where an arrow connecting two terms means that the variable at the starting point drives the evolution of that at the ending point. Next, we incorporate the role of the measurement, which has a twofold effect:  on the one hand, it enables to  acquire information, i.e., to  \emph{reduce the uncertainty} about the mechanical quadrature $\hat X_m$; this acquisition happens indirectly through the optmechanical coupling  and requires that we keep track of the stochastic component. On the other hand, the measurement \emph{introduces disturbance}, which directly affects the conjugate quadrature ($\hat X_\mathrm{in}$) and then, through the dynamics, reaches the mechanical system.

\begin{figure*}[t!]
\includegraphics[width=1\linewidth]{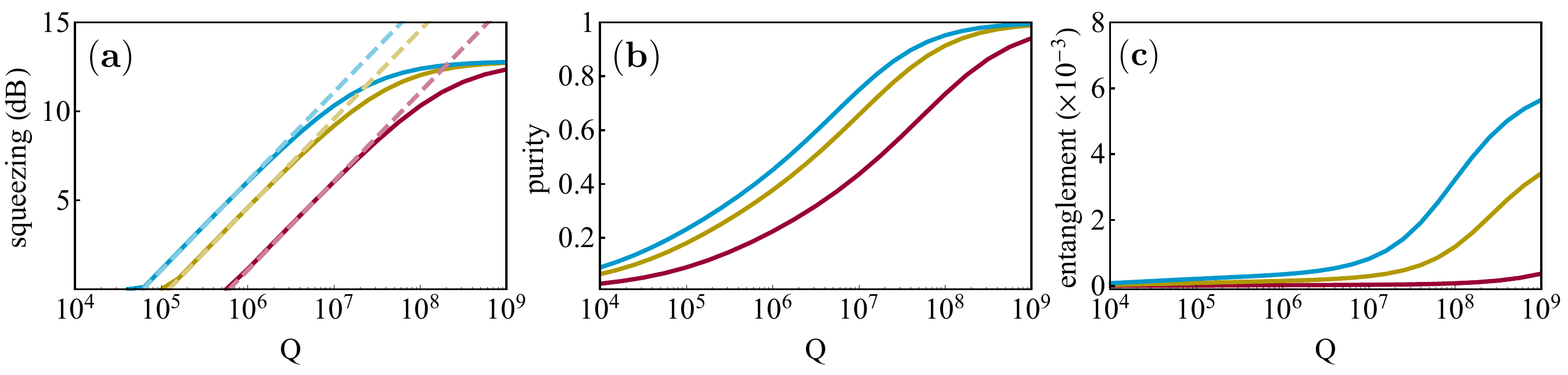}
\caption{({\bf a}) Mechanical squeezing in a stroboscopic BAE measurement. Solid lines are for the numerical solution and dashed lines are for the prediction based on 
Eqs.~\eqref{SqXextra}, \eqref{SqPextra}. The curves are for different strength, parametrized by the number of photons $N_\mathrm{p}=10^6$ (red), $N_\mathrm{p}=5\times10^6$ (yellow) and $N_\mathrm{p}=10^7$ (cyan). Other parameters are $g_0=5\times10^{-4}\omega_m,\, \bar n=1000,\,\eta=1\,,\kappa=15\omega_m,\,\tau=0.3\omega_m^{-1}$. Mechanical purity ({\bf b}) and entanglement \Me{(measured by the logarithmic negativity)} between optical cavity and mechanical resonator ({\bf c}) for the same set of parameters. \label{f:PlotComparison}}
\end{figure*}

In the ideal QND case [see Fig.~\ref{f:MeasScheme}({\bf a})] these two effects fully decouple. Fluctuations are reduced along $\hat X_m$ and increased in $\hat P_m$ (backaction heating). Graphically, this corresponds to the fact that no arrow points toward $\hat X_m$, and hence no noise can drive it. Likewise, no arrow originates from $\hat P_m$, which `absorbs' all the backaction. Thus backaction confinement  enables repeated measurements of the same quadrature with no added noise, which is the working principle of BAE measurements. When we take into account the finite mechanical evolution [cf. Fig.~\ref{f:MeasScheme}({\bf b})], the non-QND terms open new paths (dashed arrows) for both backaction and conditioning to spread, with the following consequences: information is now acquired about \emph{both} mechanical quadratures (and hence fluctuations of the conditional state are reduced in both directions) which entails  that (i) the measurement purifies the state. Similarly, measurement backaction is no longer confined to $\hat P_m$ but extends to both quadratures, i.e.,~(ii) the amount of squeezing is reduced with respect to the ideal case. Finally, information is simultaneously acquired about both the cavity and the mechanics (see multiple arrows incoming at $\hat P_c$); such joint reduction of the uncertainty implies that (iii) correlations between cavity and mechanics are built. Depending on the occupancy of the mechanical bath, this may even lead to entanglement being established between the two resonators. We want to remark that, while the limitation (ii) posed by non-QND terms is known, their beneficial effects (i) and (iii) have not been previously appreciated. A similar situation is encountered in continuous BAE measurements, where RWA solution yields conditional squeezing with low purity, and the inclusion of counter-rotating terms lower the amount of squeezing but at the same time allows for larger purity and optomechanical entanglement~\cite{Us19}.   

To check the validity of these conclusions we numerically integrate the conditional evolution of the full optomechanical system subject to stroboscopic driving and continuous homodyne detection of the output phase quadrature (see Refs.~\cite{Marco16,Us19} for details). Free mechanical evolution during each pulse is explicitly included in the simulation, i.e. we use the optomechanical interaction in Eq.~\eqref{QND2}.
In Fig.~\ref{f:PlotComparison}({\bf a}) we show the numerical squeezing in the long-time limit (averaged over one period) and compare it with the prediction of Eq.~\eqref{SqXextra}. Our simple analytical formula shows excellent agreement except for large $Q$, where it does not capture the saturation of squeezing. Such saturation confirms our expectation (ii). Indeed, for a fixed duration of the pulse, the effects of the free mechanical evolution become more prominent for larger $Q$.
From Fig.~\ref{f:PlotComparison}({\bf b}) we see that a realistic stroboscopic BAE measurement actually generates highly pure conditional squeezed states. Mixing of the two quadratures, present for any finite value of the sideband parameter, implies the \emph{simultaneous squeezing and cooling the mechanics} by stroboscopic BAE measurement, as predicted in (i). Finally, in panel ({\bf c}) the conditional optomechanical entanglement is displayed (quantified by the logarithmic negativity) which confirms (iii). Notice that entanglement is present in the high temperature regime. 

\section{Verification of the mechanical state: stroboscopic tomography}\label{s:Verification}
Essential to any conditional protocol is a verification part. While in the previous sections we have focussed on state preparation through measurement, 
here we calculate how well a quadrature can be measured in a train of pulses.
This is also known as \emph{retrodiction}~\cite{Gammelmark13,Zhang17}. State verification via retrodiction has been recently employed to verify the quantum trajectory of a continuously driven optomechanical system~\cite{Rossi19}. 
The final result of a stroboscopic measurement is a measurement value with a given confidence interval.
Repeatedly preparing and measuring a state allows for full tomography.
Since it makes sense to keep measuring until the resonator is no longer correlated with its initial state, the resonator state at the end of the measurement is again a conditionally squeezed state as discussed above, independent of the initial state.

In a stroboscopic measurement of a harmonic oscillator, its position is measured at regular intervals.
This results in a string of measurement results $\vec y=(y_0,y_1,y_2,\ldots)$,
which are correlated with the actual position at that time $y_i=x_i+m_i$.
The measurement errors $\{m_i\}$ are normally distributed as the Kraus operator corresponding to the measurement \eqref{Kraus} predicts. 
Specifically, if the oscillator is in a position eigenstate $\rho=\ket x\bra x$, the measurement probability distribution is $p(P_c)=\tr{\hat\Upsilon\dagg\hat\Upsilon\rho}\propto\exp[-(P_c/\chi-x)^2/\chi^{-2}]$. 
Thus, the $\{m_i\}$ are drawn from a Gaussian distribution of zero mean, with variance $\sigma_m^2=1/(2\chi^2)$.
In order to realize a QND measurement, the time $t_n$ between measurements has to be an integer multiple of half a period, $t_n=(n+1)\pi/\omega_m$. \Me{This is because $[\hat x(t_n),\hat x(t_m)] = 0$, i.e.  mechanical position becomes a QND observable at these stroboscopic times.}
In this regime, the problem becomes classical, as the measurement backaction is evaded.
Here we choose the time between measurements to be as short as possible, $T=\pi/\omega_m$, 
such that $\gamma T=\pi/Q$. In between each measurement, the position of the oscillator decays, due to damping, and gets a random contribution from the thermal noise acting, 
$x_i=e^{-\gamma T/2}x_{i-1}+d_i$.
The random numbers $d_i$ also follow a normal distribution of mean zero and standard deviation
$\sigma_d^2 = (\bar n+1/2) (1-e^{-\gamma T})$
which follows from the fluctuation-dissipation theorem (or equivalently from the explicit discussion in Sec.~\ref{s:CPMap}).

In \cref{app:tomography} we show that given a string of measurement results $\vec y$, the conditional probability distribution inferred from Bayes' theorem is the normal distribution
\begin{equation}
  \vec x \sim N(\vec\mu_x+\matr Q_{xx}^{-1}\vec y/\sigma_m^2,\,\matr Q_{xx}^{-1}),
  \label{eq:conditional_x_distribution}
\end{equation}
where the correlation matrix $\matr Q_{xx}$ is given by
\begin{subequations}
  \begin{align}
  	[\matr Q_{xx}]_{11} &= \frac{1}{\sigma_{x_0}^2} + \frac{1}{\sigma_m^2}+\frac{e^{-\gamma T}}{\sigma_d^2}, \\
  	[\matr Q_{xx}]_{i,i\pm1} &=-\frac{e^{-\gamma T/2}}{\sigma_d^2},\\
  	[\matr Q_{xx}]_{ii} & = \frac{1}{\sigma_m^2}+\frac{1+e^{-\gamma T}}{\sigma_d^2},\\
  	[\matr Q_{xx}]_{nn} &= \frac{1}{\sigma_m^2}+\frac{1}{\sigma_d^2}.
  \end{align}
\end{subequations}

Interestingly, as the number of measurements goes to infinity, the matrix $\matr Q_{xx}$ can be inverted analytically (see \cref{app:inverse}).
The first element of the inverse, $[\matr Q_{xx}^{-1}]_{11}$, is the variance associated with the measured value, and as one would expect it coincides with the achieved squeezing [\cref{SqApprox}]. While perhaps this could have been inferred from the results above, a very good approximation to the inverse and thus the variance can also be found for a finite number of measurements. Furthermore, this approach yields the weight each measurement value is associated with, although the \Me{general expression} is somewhat unenlightening (see \cref{app:tomography}). \Me{For the experimentally relevant case of small $\bar{n}/Q$, measurements are weighted by exponentially reducing factors with increasingly distant measurement times [see Eq.~\eqref{eq:weights}]}. Finally, we can also show that when taking into account many measurements before and after a certain point in time, i.e., using preparation and retrodiction, the associated variance is half of \cref{SqApprox}.
 
\section{Collective entanglement and cooling of two mechanical resonators}\label{s:TwoMode}
In this Section, we show how the previous results can be extended to the case of two non-degenerate mechanical resonators.
In order to do that, we consider  two mechanical resonators of frequency $\omega_{m,1}$ and $\omega_{m,2}$ coupled to a common cavity field.
Collective BAE schemes have been proposed in this configuration for continuous and two-tone driving~\cite{HammererEPR, TwoModeBAE, HammererAnnalen}.  
We introduce the mean and the relative mechanical frequency, respectively defined as 
$\omega=(\omega_{m,1}+\omega_{m,2})/2$ and $\Omega=(\omega_{m,1}-\omega_{m,2})/2$ (we assume $\omega_{m,1}>\omega_{m,2}$ without loss of generality). We also define the collective mechanical variables
\begin{equation}
\hat X_{\pm}=(\hat X_{m,1}\pm\hat X_{m,2})/\sqrt2\, ,\qquad \hat P_{\pm}=(\hat P_{m,1}\pm\hat P_{m,2})/\sqrt2\,,
\end{equation} 
that satisfy $[\hat X_{\pm},\hat P_{\pm}]=i$, $[\hat X_{\pm},\hat P_{\mp}]=0$. When the pulse is on, both mechanical resonators linearly couple to the  common cavity amplitude, giving 
\begin{align}\label{TwoToneQND}
\hat H_I(t)&=-g(t)\!\sum_{j=1,2}\!\hat X_c \!\left[\hat X_{m,j} \cos(\omega_{m,j} t)+\hat P_{m,j} \sin(\omega_{m,j} t)\right]   \\ \nonumber  
&=-\sqrt2 g(t)\hat X_c \left(\hat X\cos\omega t+\hat  Y \sin\omega t\right) \,. \label{GlobalQND}
\end{align}
For simplicity, we have considered the case of equal single-photon optomechanical couplings and in the second line we have rewritten the interaction in terms of the rotated collective quadratures 
\begin{align}
\hat X&=\hat X_+\cos\Omega t+\hat  P_- \sin\Omega t\,,\\
\hat Y&=\hat P_+\cos\Omega t-\hat  X_- \sin\Omega t\,,
\end{align}
which still form a conjugate pair $\bigl[\hat X,\hat Y\bigr]=i$. Thanks to this change of variables we see that Eq.~\eqref{GlobalQND} has the same form as~\eqref{QND2} and therefore we can rely on our previous analysis. In particular, for $g(t)=g \,\delta(t-k\pi/\omega)$ we recover the ideal case of stroboscopic QND interaction $\hat U\approx e^{i\sqrt2 \chi \hat X_c \hat X}$ (here $k=1$). This corresponds to pulsing every half of the fundamental period $2T_1T_2/(T_1+T_2)$, where $T_j$ are the single mechanical periods.  
Like in the single-mode case, we also include mechanical dissipation. For simplicity, in the following we consider equal mechanical damping rates and same occupancies for the two baths. For non-degenerate mechanical modes, these conditions may entail adjusting the local temperatures of the baths to achieve the same occupancy. 

From the discussion of Sec.~\ref{s:StroboBAE} we conclude that the stroboscopic steady state is a squeezed thermal state in the collective variables $\hat X$ and $\hat Y$, with  
the variance reduced along $\hat X$ and heated up along $\hat Y$ by the backaction. We now want to express the state in terms of the original local variables.
For this purpose, it is useful to parametrize the steady state $\sigma_{ss}$ of single mode BAE measurements [cf. Eqs.~\eqref{SqX},~\eqref{SqP}]  as 
\begin{equation}\label{CollectiveSigma}
\sigma_{ss}=\mathrm{diag}\left[(n_\mathrm{eff}+\tfrac12)e^{-r_\mathrm{eff}},(n_\mathrm{eff}+\tfrac12)e^{r_\mathrm{eff}}\right] \, ,
\end{equation}
where an explicit expression of $n_\mathrm{eff}, r_\mathrm{eff}$ can be obtained by inverting Eqs.~\eqref{SqX} and~\eqref{SqP} (one also needs to rescale $g\rightarrow g/\sqrt2$). Next we notice that the vector of original quadratures $\hat Q=(\hat X_{m,1},\hat P_{m,1},\hat X_{m,2},\hat P_{m,2})^T$ and that of collective ones $\hat Q'=(\hat X,\hat Y,\hat W,\hat Z)^T$ are related via the following transformation $\hat Q'=\hat U_{BS}^\dagger \,e^{i\Omega t(\hat b_1^\dagger \hat b_1-\hat b_2^\dagger \hat b_2)}\hat Q e^{-i\Omega t(\hat b_1^\dagger \hat b_1-\hat b_2^\dagger \hat b_2)} \hat U_{BS}$; here  $\hat Z=\hat X_-\cos\Omega t+\hat  P_+ \sin\Omega t$ and $\hat W=\hat P_-\cos\Omega t-\hat  X_+ \sin\Omega t$ are the other two collective rotated quadratures and $\hat U_{BS}$ is a beam splitter transformation. Note that the modes are rotated in opposite directions before getting mixed. By transforming state \eqref{CollectiveSigma} accordingly we get
\begin{equation}\label{TMTS}
\sigma_{ss}=
\left(
\begin{array}{cc}
A & C  \\
C^T & A
 \end{array}
\right)\,,
\end{equation}
where $A=(n_\mathrm{eff}+\frac12) \cosh 2 r_\mathrm{eff}\mathbb{1}_2$ and $C= -(n_\mathrm{eff}+\frac12) \sinh 2 r_\mathrm{eff}\sigma_z$ (with $\sigma_z$ the $z$-Pauli matrix), namely a two-mode squeezed thermal state. This state is known to be entangled if and only if $\nobreak{r_\mathrm{eff}>\ln(\sqrt{1+2n_\mathrm{eff}})}$. Therefore frequent measurements modeled by the pulses may induce entanglement between the two non-interacting resonators.
 
The same argument can be repeated for two-mode cooling, which corresponds to pulsing every quarter of the fundamental period $2T_1T_2/(T_1+T_2)$. Indeed, Eqs.~\eqref{SqXApproxCool} and \eqref{SqPApproxCool} are also in the form of a squeezed thermal state, although with very little squeezing. Therefore the two-mode stroboscopic steady state is 
still of the form \eqref{TMTS}, the difference being that now we have $n_\mathrm{eff}\approx 0$ and smaller $r_\mathrm{eff}$ (compared with the previous case). For 
both $n_\mathrm{eff}, \,r_\mathrm{eff}\approx 0$ the steady state has a large overlap with the vacuum of the two modes. Notice however that for $n_\mathrm{eff}=0$ the state is entangled for any value of $r_\mathrm{eff}>0$.

\section{Experimental considerations}\label{s:Exp}

The principal considerations for implementing stroboscopic optomechanics concern the appropriate hierarchy of time scales $\kappa^{-1}\ll\tau\ll\omega_m^{-1}$ and sufficient measurement strength $\chi$.
Low bath occupation $\bar n$ and high quality factors $Q$ facilitate access to the quantum regime. However, experimental non-idealities have to be taken into account as well.
These can include sub-unity detection efficiency, optical absorption heating, mechanical frequency drift, and spurious mechanical modes, among others.

Given the recent progress with measurement-based quantum state preparation~\cite{Rossi2018, Rossi19} with membrane-in-the-middle optomechanical systems, we discuss this platform first.
With MHz resonance frequencies, sufficiently short (sub-microsecond) pulses are readily implemented using standard modulation techniques.
Such pulses could be accommodated in short ($L\sim 1\,\mathrm{mm}$), medium-finesse resonators with $\kappa/2\pi\gtrsim 15\,\mathrm{MHz}$, and the cavity output detected with high efficiency, as previously demonstrated~\cite{Rossi2018, Rossi19}. Optical power levels tolerated in continuous-wave operation~\cite{Rossi2018} suggest $\chi= g \tau \sim 0{.}1$ can be achieved without significant device heating at a temperature of $\sim 10\,\mathrm{K}$, or $\bar n\sim 10^5$.
Since stroboscopic operation would lower the thermal load from optical absorption by a factor of order $\omega_m \tau< 1$, even higher $\chi$ may be possible, provided instabilities are avoided and the cavity lock maintained.
If the experiment were implemented with soft-clamped membrane resonators~\cite{Rossi2018, Rossi19,Tsaturyan17}, very high quality factors $Q\sim10^9$ are available
(for comparison with some of the results presented so far,  gentle laser  pre-cooling can be assumed to trade equivalent bath occupancy with quality factor,  leaving the \Me{ratio $\bar n/ Q$} constant).
With this set of parameters, significant levels of squeezing can be achieved, see Fig.~\ref{f:PlotComparison}.

In practice, however, other mechanical modes at harmonics of the stroboscopic sampling frequency  
contribute to the measured signal, in an effect known as aliasing in the context of periodically sampled data.
This would lead to spurious noise and interactions, and a degradation of the prepared state.
In contrast to pulsed optomechanics~\cite{VannerPulsed}, which offers virtually no spectral discrimination of mechanical modes at all, the stroboscopic protocols are only sensitive to spurious modes that coincide with a harmonic of the sampling frequency.
Yet soft-clampled membrane resonators with their high density of states outside the bandgap, in which the high-Q modes lie, would  be strongly affected by this effect.
Membrane-in-the-middle setups with mechanical systems that feature a sparser mode spectrum, such as trampoline resonators~\cite{ReinhardtPRX16,NortePRL16}, 
may therefore be preferable, provided sufficiently high Q-factors and/or low temperatures can be achieved.

Nanophotonic structures could be an alternative platform of interest.
Their already sparse mechanical mode spectrum could conceivably be engineered to be sufficiently anharmonic. 
Measurement strengths as high as $\chi\approx 0{.}1$ have already been demonstrated with single optical pulses much shorter ($\tau=20\,\mathrm{ns}$) than the mechanical period ($2\pi/\omega_m\approx 300\,\mathrm{ns}$)~\cite{Muhonen19}. In combination with high-efficiency readout and efficient heat removal~\cite{Qiu19arxiv}, stroboscopic optomechanics might also allow generation of squeezed and entangled states in such systems.

\section{conclusions}\label{s:Conclusions}
In this work we provided a description of the conditional dynamics of an optomechanical system driven by a train of pulses.  
We showed that the resulting framework---dubbed \emph{stroboscopic quantum optomechanics}---provides a versatile toolbox for measurement-based quantum control of optomechanical systems in the bad-cavity regime, ranging from ground state cooling to mechanical squeezing, and applicable to single as well as multimode optomechanics. Crucially, it enables the generation and characterization of measurement-based squeezing and entanglement. Stroboscopic driving alleviates the requirements of pulsed protocols based on a single (or a few) pulse(s). 

\section{Acknowledgments.} 
D.~M.~acknowledges support by the Horizon 2020 ERC Advanced Grant QUENOCOBA (grant agreement 742102).
A.~N.~acknowledges a University Research Fellowship from the Royal Society and additional support from the Winton Programme for the Physics of Sustainability.
A.~S.~acknowledges funding from the European Union's Horizon 2020 research and innovation program (European Research Council project Q-CEOM, grant agreement No 638765).
This work was supported by the European Union Horizon 2020 research and innovation programme under grant agreement No 732894 (FET Proactive HOT).

\begin{appendix}
\section{Derivation of the pulsed interaction}\label{app:Derivation}
In the following we derive the complete expression of the unitary evolution Eq.~\eqref{U} induced by the pulsed interaction and discuss when the simple QND expression Eq.~\eqref{Simple} is recovered. At a formal level, the evolution $\hat U(t,t_0)$ can be equivalently described by the  Magnus expansion 
\begin{equation}\label{UMagnus}
\hat U(t,t_0)=\exp \sum_{k=1}^\infty \hat \Omega_k(t,t_0)\,,
\end{equation} 
which comes in the form of a non-ordered exponential. Compared to Eq.~\eqref{U}, the complexity of the expression has just been shifted to the argument of the exponential. The first two terms of the Magnus expansion are given by
\begin{subequations}
\begin{align}
\hat \Omega_1(t,t_0)&=-i\int_{t_0}^t \mathrm{d}t_1\hat H_I(t_1)\,,\label{Omega1} \\
\hat \Omega_2(t,t_0)&=-\frac12\int_{t_0}^t \mathrm{d}t_1 \int_{t_0}^{t_1} \mathrm{d}t_2 [\hat H_I(t_1),\hat H_I(t_2)]\,, \label{Omega2}
\end{align}
\end{subequations}
where $\hat H_I(t)$ is the linearized interaction in the rotating frame, as given in Eq.~\eqref{QND2}. For concreteness, let us consider a rectangular pulse of length $\tau$ centered at the origin, described by the normalized profile $\varepsilon(t)=\frac{1}{\sqrt{\tau}}[\Theta(t-\tau/2)-\Theta(t-\tau/2)]$. When the pulse drives the optomechanical cavity on resonance,
the evolution of the field amplitude inside the cavity is well approximated by $\dot \alpha=-\frac{\kappa}{2}\alpha+\sqrt{\kappa N_\mathrm{p}}\varepsilon(t)$, where we neglected the mechanical response during the short interaction time. The solution reads    
\begin{equation}
\alpha(t)=2\sqrt{\frac{N_\mathrm{p}}{\kappa \tau}}\left[f_+(t)\Theta(t+\tau/2)-f_-(t)\Theta(t-\tau/2) \right]\,,
\end{equation}
where we set $f_\pm(t)=1-e^{-\frac{\kappa}{2}(t\pm\frac{\tau}{2})}$. The expression captures the build-up and the decay of the coherent field inside the cavity. The time-dependent optomechanical coupling in Eq.~\eqref{QND2} is $g(t)=g_0\alpha(t)$. Notice that in the fast cavity limit the expression reduced to a rectangular pulse of height $g_\mathrm{ad}=2g_0\sqrt{\frac{N_\mathrm{p}}{\kappa \tau}}$.

We can now compute the expressions~\eqref{Omega1},~\eqref{Omega2} for this profile.  For late times $t\gg \tau$ we get 
\begin{align}
\hat \Omega_1&=i\chi\hat X_c\left(\hat X_m + \frac{2\omega_m}{\kappa}\hat P_m\right)\,, 
&\hat \Omega_2=i\frac{\zeta}{2} \hat X_c^2\,, 
\end{align}
where the prefactors read
\begin{subequations}
\begin{align}
\chi&=2 \left(\frac{g_\mathrm{ad}}{\omega_m}\right) \frac{\sin\left(\frac{\omega_m\tau}{2}\right)}{1+4\left(\frac{\omega_m}{\kappa}\right)^2}\,, \\
\zeta&=16 \left(\frac{g_\mathrm{ad}}{\omega_m}\right)^2 \frac{\sinh^2\left(\frac{\kappa\tau}{4}\right)}{\left(\frac{\kappa}{\omega_m}\right)^3+4\left(\frac{\kappa}{\omega_m}\right)} \,.
\end{align}
\end{subequations}
In the above expression we set $g_\mathrm{ad}=2g_0\sqrt{\frac{N_\mathrm{p}}{\kappa \tau}}$. In the adiabatic limit $\chi$ reduces to $\chi_\mathrm{ad}=g_\mathrm{ad}\tau\equiv 2 g_0 \sqrt{\frac{N_\mathrm{p}\tau}{\kappa}}$, which is the expression given in the main text. Also notice that, as expected for the problem at hand, all the nested commutators corresponding to higher order $\hat \Omega_{k\ge3}$ identically vanish. The evolution thus takes the following exact expression 
\begin{equation}\label{UnonQND}
\hat U=e^{i\tfrac{\zeta}{2} \hat X_c^2} e^{i\chi\hat X_c\hat X_m  + i\chi\left(\tfrac{2\omega_m}{\kappa}\right)\hat X_c\hat P_m} \, .
\end{equation}
Two extra terms have appeared compared to Eq.~\eqref{Simple}. A single-mode operator that is responsible for squeezing of the cavity field
and an interaction term that spoils the QND character of the quadrature $\hat X_m$. Note that both terms are present for any length of the pulse, even though the spurious term is suppressed by a factor $2\omega_m/\kappa$. The QND limit can then be recovered only for a vanishing sideband parameter $\omega_m/\kappa$. 

\section{Complete expression for stroboscopic cooling}\label{app:Cooling}
The complete expressions of the mechanical variances at the stroboscopic steady state in case of cooling [Eqs.~\eqref{SqXApproxCool} and~\eqref{SqPApproxCool}], up to $\mathcal{O}(Q^{-1})$, are given by 
\begin{widetext}
\begin{align}
\sigma_{X_m}&=\frac{\sqrt{4+\chi^4}-\chi^2}{4}+ \frac{\pi  \left[-4 \bar n \chi ^2+\frac{4 \bar n \left(\chi ^4+2\right)+\chi ^6+2 \chi ^4+4 \chi ^2+4}{\sqrt{\chi ^4+4}}-\chi ^4-2 \chi ^2-2\right]}{8 Q \chi ^2}\,,  \\
\sigma_{P_m}&=\frac{\sqrt{4+\chi^4}+\chi^2}{4}+\frac{\pi  \left[2 \bar n \left(\chi ^4+2\right)+\chi ^4-\sqrt{\chi ^4+4}+2\right]}{4 Q \chi ^2 \sqrt{\chi ^4+4}}\,. 
\end{align}
\end{widetext}

\section{Extra-cavity stroboscopic measurement}\label{app:ExtraMeas}
In this appendix we provide more details about the derivation of expressions Eqs.~\eqref{SqXextra}, ~\eqref{SqPextra}. 
Our starting point will be Eq.~\eqref{HExtended}, where for simplicity we assume the ideal QND interaction given in Eq.~\eqref{Simple}. 
As described in Sec.~\ref{s:Improved}, we consider delta-correlated quantum noise $[\hat a_{\mathrm{in},t},\hat a_{\mathrm{in},t'}^\dag]=\delta(t-t')$. 
The operators $\hat a_{\mathrm{in},t}$ are singular and have dimension $(\mathrm{time})^{-1/2}$; Hermitian combinations of them thus cannot be directly associated with observables to be measured. To remedy that,  
it is customary to introduce the so-called quantum Wiener increment, defined as $\Delta \hat W_{\mathrm{in}}= \int_{t}^{t+\tau}\mathrm{d}t'\hat a_{\mathrm{in},t'}$. For a short interval of time we have
$d \hat W_{\mathrm{in}}= \hat a_{\mathrm{in},t}\mathrm{d}t$ (obtained for $\tau\rightarrow\mathrm{d}t$). 
From this expression one can define proper dimensionless modes $\hat A_{\mathrm{in}}$ via $d \hat W_{\mathrm{in}}=\hat A_{\mathrm{in}}\sqrt{\tau}$, which are non-singular $[\hat A_{\mathrm{in}},\hat A_{\mathrm{in}}^\dag]=1$ and can be associated with a measurement. The corresponding quadrature operators are $\hat X_\mathrm{in}=(\hat A_{\mathrm{in}}+\hat A_{\mathrm{in}}^\dag)/\sqrt2$, $\hat P_\mathrm{in}=i(\hat A_{\mathrm{in}}^\dag-\hat A_{\mathrm{in}})/\sqrt2$ and are associated to quadrature measurements of the outgoing light field. Since the expression in Eq.~\eqref{UExtended} still contains interaction which are bilinear, we can repeat the same analysis of Sec.~\ref{s:Model} to get the conditional covariance matrix of the optomechanical system, now on the enlarged phase space with coordinate $(X_\mathrm{in},P_\mathrm{in},X_c,P_c,X_m,P_m)$. 

\section{Stroboscopic Tomography}\label{app:tomography}
\subsection{\Me{General derivation}}
We would like to model a series of BAE measurements on a damped harmonic oscillator in a thermal environment.
In order to do so, we need to find the probability distribution of $\vec x$ given some measurement record $\vec y$, $P(\vec x|\vec y)$, assuming that  the parameters $\sigma_m, \sigma_d$, $\gamma$ are known.

From the description in the main text, we can determine the conditional probability distribution for the measurement outcomes
\begin{equation}
	P(\vec y|\vec x) = 
	\prod_{i=0}^{N-1}\frac{1}{\sqrt{2\pi\sigma_m^2}}\exp\left(-\frac{(y_i-x_i)^2}{2\sigma_m^2}\right).
  \label{eq:conditional_probability}
\end{equation}
as well as our prior
\begin{equation}
  \begin{aligned}
  	P(\vec x)&= \frac{1}{\sqrt{2\pi\sigma_{x_0}^2}}\exp\left(-\frac{x_0^2}{2\sigma_{x_0}^2}\right)\\
  	&\times\prod_{i=1}^{N-1}\frac{1}{\sqrt{2\pi\sigma_d^2}}
  	\exp\left(-\frac{(x_i-e^{-\gamma T/2}x_{i-1})^2}{2\sigma_d^2}\right).
  \end{aligned}
  \label{eq:prior}
\end{equation}
The initial variance for $x_0$ could be from a thermal state. 
Optionally, we could already have performed measurements at that point, in which case the initial variance and mean are given by the resulting state.
We can thus write down the joint probability distribution
$P(\vec x, \vec y)=P(\vec y|\vec x)P(\vec x)$.
\begin{equation}
  P(\vec x,\vec y)=P(\vec x)P(\vec y|\vec x)\propto \exp\left(-\frac12 (\vec x^\top,\vec y^\top)\matr Q\begin{pmatrix}\vec x\\\vec y\end{pmatrix}\right),
  \label{eq:joint_prob_dist}
\end{equation}
where $\matr Q$ has entries
$\matr Q_{yy} = -\matr Q_{xy} = (1/\sigma_m^2)\id$, and
\begin{equation}
  \begin{aligned}
  	&[\matr Q_{xx}]_{ij} = -\frac{e^{-\gamma T/2}}{\sigma_d^2}(\delta_{i,j+1}+\delta_{i,j-1})\\
  	&+\left[ \frac{1}{\sigma_m^2} +\frac{1+e^{-\gamma T}}{\sigma_d^2}-
  \delta_{i,N}\frac{e^{-\gamma T}}{\sigma_d^2}
  + \delta_{i,1}\left( \frac{1}{\sigma_{x_0}^2}-\frac{1}{\sigma_d^2}\right)
   \right]\delta_{i,j}.
  \label{eq:Qxx}
  \end{aligned}
\end{equation}

$P(\vec x,\vec y)$ is a normal distribution 
 $(\vec x,\vec y)\sim N\left[ \vec \mu,\matr\Sigma \right]$,
 with mean $\vec\mu=0$ and covariance matrix $\matr\Sigma=\matr Q^{-1}$, which can be found via block matrix inversion
\begin{equation}
  \matr\Sigma=
  \mat{\matr\Sigma_{xx} & \matr\Sigma_{xy} \\ \matr\Sigma_{yx} & \matr \Sigma_{yy}}
\end{equation}
where
\begin{subequations}
  \begin{align}
  	\matr \Sigma_{yy}&=\sigma_m^2+\matr\Sigma_{xx},\\
  	\matr\Sigma_{xx}&=(\matr Q_{xx}-1/\sigma_m^2)^{-1}=\matr\Sigma_{xy}=\matr\Sigma_{yx}.
  \end{align}
  \label{eq:Sigmas}
\end{subequations}

Given this, the conditional distribution for $\vec x$ can be derived from the joint distribution
\begin{equation}
  \vec x \sim N(\matr\Sigma_{xy}\matr\Sigma_{yy}^{-1}\vec y,
  \matr\Sigma_{xx}-\matr\Sigma_{xy}\matr\Sigma_{yy}^{-1}\matr\Sigma_{xy}).
  \label{eq:conditional_dist_x}
\end{equation}

For now, we are only interested in the mean and variance of the first entry. 
We thus need to determine
\begin{equation}
  \matr\Sigma_{xx}\matr\Sigma_{yy}^{-1}
  =\left( \sigma_m^2\matr \Sigma_{xx}^{-1}+\id \right)^{-1}
  =\matr Q_{xx}^{-1}/\sigma_m^2.
  \label{eq:estimate_mean}
\end{equation}
and 
\begin{equation}
  \matr \Sigma_{xx}-\matr\Sigma_{xy}\matr \Sigma_{yy}^{-1}\matr\Sigma_{xy}
  =\matr\Sigma_{xx}\left( 1-\sigma_m^{-2}\matr Q_{xx}^{-1} \right)=\matr Q_{xx}^{-1}.
  \label{eq:estimate_variance}
\end{equation}
Note that $\matr\Sigma_{xx}$ commutes with the matrix in round brackets.
\Me{The best estimates for the positions $\vec{x}$ can therefore be obtained from the measurement results by multiplying the latter with the weights \eqref{eq:estimate_mean} and their covariance is given through  \eqref{eq:estimate_variance}. In both cases we need to determine $\matr Q_{xx}^{-1}$, which is done in the following section.}

\subsection{\Me{Explicit calculation of the inverse of the matrix $\matr Q_{xx}$}}\label{app:inverse}
\Me{In order to compute the inverse it is useful to consider the matrix}  
\begin{equation}
  	\matr M = \mat{1 & a & 0 & \cdots & 0 & 0 & 0 \\
  	  a & b & a & \cdots & 0&0&0\\
  	  0&a&b&\cdots&0&0&0\\
  	  \vdots&\vdots&\vdots&\ddots&\vdots&\vdots&\vdots\\
  	  0&0&0&\cdots&b&a&0\\
  	  0&0&0&\cdots&a&b&a\\
  	0&0&0&\cdots&0&a&1}.
  	\label{eq:tridiagonal_matrix}
\end{equation}
\Me{We show in Appendix~\ref{app:exp_case} that for experimentally relevant parameters this coincides with a rescaled version of the matrix $\matr Q_{xx}$, namely $\matr M=\matr Q_{xx}/[\matr Q_{xx}]_{11}$.}
The entries of the inverted matrix are~\cite{Usmani1994}
\begin{equation}
  M^{-1}_{ij} =
  (-1)^{i+j}a^{|i-j|}
  \begin{cases}
  	\theta_{i-1}\phi_{j+1}/\theta_n & \text{if }i\leq j\\
  	\theta_{j-1}\phi_{i+1}/\theta_n & \text{if }i\geq j,
  \end{cases}
  \label{eq:general_inverse}
\end{equation}
where $\theta$ and $\phi$ fulfil certain recurrence relations. 
In our case, they are in fact the same, and we have
\begin{equation}
  \begin{aligned}
  	\theta_i &= b \theta_{i-1} - a^2 \theta_{i-2},\qquad \theta_0=\theta_1=1,\\
  	\theta_n &=   \theta_{n-1} - a^2 \theta_{n-2},\qquad \phi_i=\theta_{n+1-i}.
  \end{aligned}
  \label{eq:recurrence_relation}
\end{equation}
The solution can be found from the characteristic polynomial $x^2-bx+a^2=0$, which has roots
\begin{equation}
  \xi_\pm = \frac{b}{2}\pm\sqrt{\frac{b^2}{4}-a^2}.
  \label{eq:roots}
\end{equation}
Fitting the general solution to the boundary conditions, we find
\begin{equation}
  \theta_i = \frac{1}{\sqrt{b^2-4a^2}}\left[ \xi_-^i(\xi_+-1)-\xi_+^i(\xi_--1) \right],
  \label{eq:theta_i}
\end{equation}
but with $\theta_n$ determined by \cref{eq:recurrence_relation} above.
This allows us to write the inverse of the above matrix in an exact analytical,
albeit unwieldy manner
\begin{equation}
  \begin{aligned}
  	(\matr M^{-1})_{ij}&= \frac{(-1)^{i+j}a^{|i-j|}}{\sqrt{b^2-4a^2}}
  	\frac{\nu(i-1)\nu(n-j)}{\nu(n-1)-a^2\nu(n-2)},\\
  	\nu(i)&\equiv\xi_-^i(\xi_+-1)-\xi_+^i(\xi_--1).
  \end{aligned}
  \label{eq:analytical_inverse}
\end{equation}
This exact formula for the inverse represents the central result of this appendix.
We consider simplifications that arise in certain cases below.

\subsubsection{Physically relevant case}\label{app:relevant_case}
The relevant case is $b^2/4>a^2$. In this case $\xi_+>\xi_-$, such that in the limit of a large number of measurements $n\to\infty$, the formula for the inverse turns into
\begin{equation}
  \begin{cases}
  (\matr M^{-1})_{ij} = \frac{(-1)^{i+j}a^{|i-j|}}{\sqrt{b^2-4a^2}}
  \frac{\nu(i-1)}{\xi_+^{j-1}-a^2\xi_+^{j-2}}, & i<j,\\
  \matr M_{ji}=\matr M_{ij}, & i>j.
  \end{cases}
  \label{eq:case1}
\end{equation}
It is only exact for $n\to \infty$, and is a good approximation if $(a\xi_-/\xi_+)^n$ is small.
The weights for the measurement of the initial state are the special case $i=1$, i.e.,
\begin{equation}
  (\matr M^{-1})_{1j} \simeq
  \frac{(-a)^{|1-j|}}{\xi_+^{j-1}-a^2\xi_+^{j-2}}.
  \label{eq:weights}
\end{equation}
Another useful special case is $i=j$, in which case \cref{eq:case1} simplifies to
\begin{equation}
  (\matr M^{-1})_{ii} = \frac{(\xi_-/\xi_+)^{i-1}(\xi_+-1)-\xi_-+1}{\sqrt{b^2-4a^2}(1-a^2/\xi_+)}.
  \label{eq:variance_formula}
\end{equation}

On the other hand, the variance in steady-state ($i\to\infty$, but $n/i\gg1$), which corresponds to the variance when taking all measurements before and after a specific point in time into account, 
\begin{equation}
  \lim_{n\to\infty}(\matr M^{-1})_{n/2,n/2} = \frac{1-\xi_-}{\sqrt{b^2-4a^2}(1-a^2/\xi_+)}.
  \label{eq:steadystate_variance_formula}
\end{equation}

\subsubsection{Other cases}
For completeness, we mention the other case is $b^2/4<a^2$, which implies $\xi_-=\xi_+^*\equiv \xi$, such that
\begin{equation}
  (\matr M^{-1})_{ij} = \frac{(-1)^{i+j}a^{|i-j|}}{\sqrt{b^2-4a^2}}
  \frac{2\Im[\xi^{i-1}(\xi^*-1)]\Im[\xi^{n-j}(\xi^*-1)]}{i\Im[\xi^{n-2}(\xi^*-1)-a^2\xi^{n-1}(\xi^*-1)]},
  \label{eq:case2}
\end{equation}
where $\Im[x]$ denotes the imaginary part of $x$.

Finally, if $b^2/4=a^2$, the matrix is not invertible.

\subsection{\Me{Variance and measurement weights for experimentally relevant parameters}}\label{app:exp_case}
The actual matrix we are interested in has parameters
\begin{subequations}
  \begin{align}
  	[\matr Q_{xx}]_{11} &= \frac{1}{\sigma_{x_0}^2} + \frac{1}{\sigma_m^2}+\frac{e^{-\gamma T}}{\sigma_d^2}, \\
  	[\matr Q_{xx}]_{i,i+1} &=-\frac{e^{-\gamma T/2}}{\sigma_d^2},\\
  	[\matr Q_{xx}]_{ii} & = \frac{1}{\sigma_m^2}+\frac{1+e^{-\gamma T}}{\sigma_d^2},\\
  	[\matr Q_{xx}]_{nn} &= \frac{1}{\sigma_m^2}+\frac{1}{\sigma_d^2}.
  \end{align}
\end{subequations}

In order to use the analytical matrix inverse derived in \cref{app:inverse}, we define $\matr M=\matr Q_{xx}/[\matr Q_{xx}]_{11}$, which has $\matr M_{11}=1$, $\matr M_{nn}\approx1$, $\matr M_{i,i+1}=a=[\matr Q_{xx}]_{i,i+1}/[\matr Q_{xx}]_{11}$ and $\matr M_{ii}=b=[\matr Q_{xx}]_{ii}/[\matr Q_{xx}]_{11}$.
The fact that the last element of the diagonal of $\matr M$ is not 1 is irrelevant if the number of measurements $n$ is large.
We can therefore take it to be 1 for simplicity.
Technically, the matrix is still invertible without this assumption, but it leads to cumbersome formulae that are not very enlightening.

For stroboscopic measurements to make sense, we require $\sigma_d^2\ll1$, \emph{i.e.}, the state is coherent for several periods.
As a result, the term with $1/\sigma_d^2$ dominates all the elements of $\matr Q_{xx}$. 
Physically, this means that the value of $x_i$ is most strongly constrained by its neighbours $x_{i-1}$ and $x_{i+1}$, and much less by the measurement or our initial guess.
This is precisely the regime of a slowly decoherering and weakly measured oscillator that we consider here. 
In this limit, 
\begin{equation}
  b^2/4-a^2 = \matr Q_{xx,11}^{-2}\left[ \frac{1}{4\sigma_m^2}+\frac{1+e^{-\gamma T}}{2\sigma_m^2\sigma_d^2}-\frac{e^{-\gamma T}}{\sigma_d^4} \right]>0,
  \label{eq:discriminant}
\end{equation}
such that we may use the formulae from \cref{app:relevant_case}.

To leading order in $\sigma_d$, $\matr M$ is the discrete Laplace operator, with $a=1$ and $b=2$, which is not invertible, so we have to go to next order to get physical answers.
Note that $\gamma T = \sigma_d^2/(n_{\mathrm{th}}+1/2)$, such that
\begin{subequations}
  \begin{align}
  	a &=\frac{[\matr Q_{xx}]_{i,i+1}}{[\matr Q_{xx}]_{11}} \simeq -1 + \sigma_d^2\left(\frac{1}{\sigma_m^2}+\frac{1}{\sigma_{x_0}^2}-\frac{1}{2n_{\mathrm{th}}+1}\right),\\
  	b &=\frac{[\matr Q_{xx}]_{ii}}{[\matr Q_{xx}]_{11}} \simeq  2 + \sigma_d^2\left(\frac{2}{2n_{\mathrm{th}}+1}-\frac1{\sigma_m^2}-\frac{2}{\sigma_{x_0}^2}\right).
  \end{align}
\end{subequations}

Using the formulae for the inverse of the matrix derived above, we can now calculate
the variance of our measurement of the initial state ($i=1$)
\begin{equation}
  Q_{xx,11}^{-1}(\matr M^{-1})_{11} = \sqrt{\frac{\pi(1/2+n_{\mathrm{th}})}{2Q\chi^2}} + \mathcal O(Q^{-3/2}),
  \label{eq:initial_variance_formula}
\end{equation}
where we have used that
\begin{equation}
  \sigma_m^2=1/(2\chi^2)
  \label{eq:measurement_variance}
\end{equation}
for the measurement we consider. \Me{This expression coincides with the amount of squeezing predicted in Eq.~\eqref{SqApprox}.}

On the other hand, the variance in steady-state ($i\to\infty$, but $n/i\gg1$), \cref{eq:steadystate_variance_formula}, simplifies to
\begin{equation}
  \lim_{n\to\infty}(\matr M^{-1})_{n/2,n/2}
  = \sqrt{\frac{\pi(1/2+n_{\mathrm{th}})}{8Q\chi^2}} + \mathcal O(Q^{-3/2}),
  \label{eq:steadystate_variance2}
\end{equation}
i.e., to leading order it is just half of \cref{eq:initial_variance_formula}.

\end{appendix}

\bibliography{MyRefs}

\end{document}